\documentstyle[11pt,aas2pp4,graphicx]{article}

\slugcomment{}

\lefthead{Hachisuka et al.}
\righthead{Distance of W3(OH)}

\begin{document}

\def\kms {km~s$^{-1}$}

\title{Water maser motions in W3(OH) and a determination of its distance}

   \author{K. Hachisuka\altaffilmark{1,2},
           A. Brunthaler\altaffilmark{3,1},
           K. M. Menten\altaffilmark{1},
           M. J. Reid\altaffilmark{4}, \\
	   H. Imai\altaffilmark{5},
	   Y. Hagiwara\altaffilmark{6,7}, \\
           M. Miyoshi\altaffilmark{6},
	   S. Horiuchi\altaffilmark{8} and
           T. Sasao\altaffilmark{9}\\
	             }

\altaffiltext{1}{Max-Planck-Institut f\"ur Radioastronomie,
                  Auf dem H\"ugel 69, 53121 Bonn, Germany}
\altaffiltext{2}{Departament de Astronom\'ia, Universitat de Val\'encia,
              Burjassot, Val\'encia 46100, Spain}
\altaffiltext{3}{Joint Institute for VLBI in Europe, Postbus 2,
              7990 AA Dwingeloo, The Netherlands}
\altaffiltext{4}{Harvard Smithsonian Center for Astrophysics,
              60 Garden Street, Cambridge, MA 02138, USA}
\altaffiltext{5}{Department of Physics, Kagoshima University, Kagoshima
              890-0065, Japan}
\altaffiltext{6}{National Astronomical Observatory, Mitaka, Tokyo 181-8588,
                 Japan}
\altaffiltext{7}{ASTRON, Westerbork Observatory, P.O. Box 2,
              7990 AA Dwingeloo, The Netherlands}
\altaffiltext{8}{Centre for Astrophysics and Supercomputing, Swinburne
              University of Technology, P.O. Box 218, Hawthorn, VIC 3122,
              Australia}
\altaffiltext{9}{Department of Space Survey and Information Technology,
              Ajou University, Suwon, 442-749, Republic of Korea}

\begin{abstract}
We report phase-referencing VLBA observations of H$_{2}$O masers near the
star-forming region W3(OH) to measure their parallax and absolute proper 
motions.
The measured annual parallax is 0.489 $\pm$ 0.017  milli-arcseconds
(2.04 $\pm$ 0.07 kpc), where the error is dominated by a systematic atmospheric 
contribution. This distance is consistent with
photometric distances from previous observations and
with the distance determined from CH$_3$OH maser
astrometry presented in a related paper.
We also find that the source driving the H$_{2}$O outflow, the
``TW-object'', moves with a 3-dimensional velocity of $>$ 7 km s$^{-1}$
relative to the ultracompact \ion{H}{2} region W3(OH).
\end{abstract}

\keywords{astrometry -- stars: individual (W3(OH)) -- masers --
stars: distances -- stellar dynamics -- stars: formation}

\section{Introduction}

The annual parallax is the most direct measurement of distances in astronomy.
The Hipparcos satellite successfully measured the distances to numerous stars 
in the Solar neighborhood, typically achieving 10\%
accuracies for distances of $\approx 100$ pc, which 
contributed significantly to many fields of modern astronomy
(e.g. Perryman et al. 1995). However, annual parallax measurements for stars
with kpc distances require sub-milliarcsecond accuracy, which has not been
achieved optically.

Very Long Baseline Interferometry (VLBI) provides the highest resolution in
astronomy. In phase-referencing VLBI, the position of a target source is
measured relative to a nearby positional reference source
(see e.g. Beasley \& Conway 1995; Ros 2003).
The feasibility of annual parallax measurements with the Very Long Baseline
Array (VLBA) has been demonstrated at low frequencies by Brisken et al. (2002) 
who measured annual parallaxes of pulsars in the Galaxy and by van Langevelde 
et al. (2000) and  Vlemmings et al. (2002) who measured distances of Galactic
OH masers associated with late type stars. 
Chatterjee et al. (2004) measured pulsar parallaxes at 5 GHz
and showed that the accuracy of astrometric
measurements improves with higher frequencies. Their results indicate that one 
can measure distances of up to a few kpc with better than 10\% uncertainty  
with VLBA astrometry of maser sources. Indeed, Kurayama et al. (2005) used the 
VLBA to measure the annual parallax of the Mira-Type star UX Cygni with high 
accuracy.

Hence, VLBA measurements allow sources spread over a large part of the 
Milky Way to have accurate parallaxes. This enables us to probe Galactic 
structure and dynamics since maser sources are spread over 
the whole Galaxy and, especially water vapor (H$_2$O) maser sources
are even found in its outer reaches (e.g. Wouterloot et al. 1993).

The 22.2 GHz transition of  H$_2$O is the most widespread
and luminous known maser line. In our Galaxy it has been detected
toward numerous evolved red giant stars and high- and low-mass star-forming
regions (see, e.g., Valdettaro et al. 2001).

W3(OH) is a region containing several high- and
intermediate-mass young stars and proto-stars of different evolutionary stages
(e.g. Wilner et al. 1999 ; Wyrowski et al. 1997, 1999).
In addition to strong OH and CH$_3$OH masers, which are seen projected
on the archetypical ultracompact \ion{H}{2} (UC\ion{H}{2}) region,
very strong H$_2$O maser emission is found toward 
the Turner-Welch (TW) Object (Turner \&\ Welch 1984; Reid et al. 1995;
Wilner et al. 1999), a protostar projected
$\approx 10^{4}$ AU east of the UC\ion{H}{2} region.
The W3(OH) H$_2$O masers were amongst the first studied with VLBI
(Moran et al. 1973). VLBI Maps of the H$_2$O maser emission have been
reported by Alcolea et al (1992).

We observed W3(OH) to measure its annual parallax and to study the internal 
dynamics of the known bipolar H$_2$O outflow from the TW object.
Moreover, our observations constitute a trial parallax and proper motion 
observation to explore the potential of utilizing H$_2$O masers as probes of 
Galactic structure. Here we report VLBA observations of the W3(OH) H$_2$O 
masers which yielded an extremely accurate parallax. 

\section{Observations and data reduction}

We observed the W3(OH) H$_{2}$O masers seven times with the NRAO VLBA 
\footnote{The VLBA is operated by the National Radio Astronomy Observatory 
(NRAO). The National Radio Astronomy Observatory is a facility of the
National Science Foundation operated under cooperative agreement by Associated
Universities, Inc.} between January 2001 and May 2002 
(see Table~\ref{obs}).
Each observation was carried out over a 4-hour period including calibrator
observations. The separations between the epochs were between two and four
months. We observed two 16-MHz bands with one band centered on the maser
velocity. The data were correlated with 1024 spectral channels in each band
with an integration time of 2 seconds. The resulting velocity channel spacing 
was 0.224 km s$^{-1}$, and we covered a velocity range of 229 km s$^{-1}$.
\begin{table}[htbp]
\caption[]{Summary of the VLBA observations. \label{obs}}
\centering
\begin{footnotesize}
\begin{tabular}{ccc}
\hline
\multicolumn{2}{c}{Epoch} & Stations \\ \hline
2001/01/28&01:11:00 -- 05:13:00 (UT)& 7 \\
2001/05/12&18:20:00 -- 22:20:00 (UT)& 9 \\
2001/07/12&14:20:00 -- 18:20:00 (UT)& 8 \\
2001/08/25&11:27:00 -- 15:27:00 (UT)& 9 \\
2001/10/23&07:35:00 -- 11:35:00 (UT)& 9 \\
2002/01/12&02:16:00 -- 06:16:00 (UT)& 9 \\
2002/05/06&18:44:00 -- 22:44:00 (UT)& 7 \\
\hline
\end{tabular}
\end{footnotesize}
\end{table}

We used ICRF 0244+624 as a phase-reference source. Its angular separation
from W3(OH) is 2.2$^\circ$. Since this source is extragalactic with a redshift
of 0.0438 (Margon \& Kwitter 1978), its  proper motion should be negligible.
The source was detected at all epochs with peak flux densities $>0.9$~Jy. 
The source was compact and unresolved as in previous VLBI observations at 
lower frequencies (Fey \& Charlot. 2000), making it an excellent 
phase-reference source.  
Typical source elevations varied from 46 to 62 degrees. We switched every 20
seconds between the sources W3(OH) and ICRF 0244+624, yielding typical
on-source times of 7 seconds.
The strong source NRAO150 was observed 
for 5 minutes every 44 minutes for delay and bandpass calibration.

The data were calibrated and imaged with standard techniques using the
NRAO Astronomical Image Processing System (AIPS) software package.
Amplitude calibration used system temperature measurements and standard
gain curves. A fringe fit was performed on ICRF 0244+624 and the solutions
were applied to W3(OH). The Kitt Peak and Los Alamos antennas did not
observe in the first epoch because of heavy snow, the Pie Town antenna was
flagged in the third and seventh epoch since most of the data was lost
because of system failures.  Also, the Saint Croix antenna was flagged in all
epochs as it produced little useful data, probably owing to its moist,
low-latitude, low-elevation site and the absence of short baselines to
this telescope.

Given an accurate geometric model in the VLBA correlator, the largest errors 
in  phase-referencing observations are introduced by a zenith delay error in 
the atmospheric model of the correlator (see Reid et al. 1999). These errors
degrade the image quality and the astrometric accuracy. The main 
contributions to the fringe-phase of the target source, after 
phase-referencing, are from a position offset and the  
atmospheric/ionospheric delay error (if the source structure is 
negligible). Because of the different behavior of the two contributions, it is 
possible to separate both effects and to estimate the position offset as well 
as a zenith delay error. We fit a model phase owing to a position
offset and a zenith delay error at each station to the calibrated phase data 
of an isolated and strong maser feature. The phase errors caused by
the zenith delay errors can then be corrected by the AIPS task CLCOR.
This correction improves the quality of the phase-referenced images and
the astrometric accuracy (Reid et al. 1999; Reid \& Brunthaler 2004; 
Brunthaler et al. 2005). The data from the Mauna Kea antenna at the 7th  
epoch was flagged, since we could not accurately estimate the zenith delay 
error of this station. After applying these corrections to the UV data
of W3(OH) we created images of the maser features for each epoch.

We determined the position of each maser spot by fitting a two-dimensional
Gaussian brightness distribution to the maps, using the AIPS task JMFIT.
The {\it formal} error of the positions were calculated from the 
signal-to-noise ratio of the fitted peak flux densities and sizes of the 
components and were typically 10 $\mu$as in right ascension  and 20 $\mu$as 
in declination.

\section{Results}

\begin{figure}
\resizebox{\hsize}{!}{\includegraphics[angle=0]{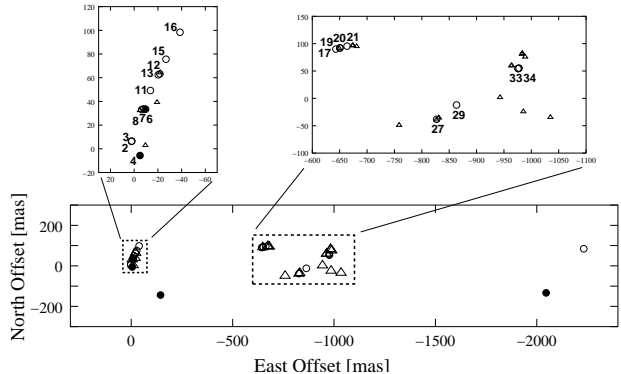}}
\caption{H$_{2}$O maser distribution. Filled circles show features detected
in 7 epochs, open circles features detected in 6 or 5 epochs detectable,
and open triangles show features detected in 4 or 3 epoch.
The numbers denote the components listed in Table 1 for which astrometrical 
fits were obtained.
}
\label{maser-distribution}
\end{figure}

We found a total of 42 distinct H$_{2}$O maser features that were
detectable over 3 epochs, usually in 2 to 10 adjacent velocity
channels with peak flux densities from a few hundred mJy to a few hundred Jy
(Tab.~\ref{maser-position}).
The masers were distributed over an area of 2.5'' $\times$ 0.5''
(Fig.~\ref{maser-distribution}), which is consistent with previous VLBI
observations (Alcolea et al. 1992).

The absolute proper motion of a
maser feature is the sum of the outflow motion in W3(OH), annual parallax,
Galactic rotation, Solar motion and peculiar motion of W3(OH).
We assumed all motions except the annual parallax to be linear.
All motions, except the internal motions, are equal for all maser features.
It can be challenging to trace exactly the same maser feature, since H$_{2}$O
masers are highly time variable and their absolute proper motions relative
to the extragalactic reference source are non-linear because of the effect
of the parallax.  For the feature identification we used not the absolute 
proper motions but the motions relative to a reference feature (feature 1 in 
Table 2) and then looked for rectilinear motions of a reasonable magnitude 
(i.e., $<100$~\kms) in each spectral channel. The relative motions in Table 2
were obtained after a phase self-calibration on the reference feature to 
reduce systematic errors.

\begin{footnotesize}
\begin{deluxetable}{lrrcrrrrccrr}
\tablewidth{33pc}
\tablecaption{Detected H$_{2}$O masers \label{maser-position}}
\tablehead{
\colhead{No.} & \multicolumn{2}{c}{Offset\tablenotemark{a}}
&\colhead{} &  \multicolumn{4}{c}{Relative proper motion\tablenotemark{b}} &
\multicolumn{2}{c}{V$_{\mbox{\scriptsize LSR}}$\tablenotemark{c}}
& \multicolumn{2}{c}{Peak flux\tablenotemark{d}} \\
\colhead{} &  \colhead{E-W} & \colhead{N-S} & \colhead{epochs} &
\multicolumn{2}{c}{E-W} & \multicolumn{2}{c}{N-S} &
\colhead{Max} & \colhead{Min} & \colhead{Max} & \colhead{Min}\\
\colhead{} & \multicolumn{2}{c}{(mas)}&\colhead{} &
\multicolumn{4}{c}{(mas~yr$^{-1}$)} &
\multicolumn{2}{c}{(km~s$^{-1}$)} &\multicolumn{2}{c}{(Jy/beam)}
}
\startdata
 1& -144.49& -144.29& 7& 0&$\pm$0&
0&$\pm$0& -50.9& -51.3& 304.6& 57.7 \nl
 2&     2.33&    6.27& 6&  0.62& 0.13&  2.30& 0.18& -47.9& -48.8&  95.8& 12.4 \nl
 3&     1.97&    6.52& 5&  0.54& 0.28&  2.63& 0.19& -47.9& -48.2& 165.5& 33.3 \nl
 4&    -4.92&   -5.68& 7&  1.06& 0.06&  1.26& 0.09& -48.2& -48.8& 335.5& 65.4 \nl
 5&    -9.45&    3.00& 4&  0.97& 0.14&  1.41& 0.19& -48.2& -48.6& 253.1& 21.6 \nl
 6&    -9.63&   33.42& 7&  0.37& 0.06&  2.73& 0.05& -48.4& -49.0& 367.0& 40.5 \nl
 7&    -8.22&   33.69& 6&  0.87& 0.11&  2.57& 0.11& -48.6& -49.6& 127.0& 12.6 \nl
 8&    -6.38&   33.18& 5& -0.15& 0.11&  3.06& 0.07& -49.8& -49.8&  12.9&  5.5 \nl
 9&    -4.83&   32.42& 4&  0.66& 0.58&  2.88& 0.18& -49.6& -49.8&  20.6&  4.9 \nl
 10&   -19.26&   39.29& 4& -0.17& 0.17&  2.81& 0.13& -48.8& -49.4&  72.3&  9.9 \nl
 11&   -13.67&   49.22& 5&  0.83& 0.08&  1.51& 0.20& -49.2& -49.6& 255.2& 21.9 \nl
 12&   -21.64&   63.22& 6&  1.09& 0.29&  1.02& 0.18& -49.0& -49.2& 126.1& 51.9 \nl
 13&   -20.30&   62.46& 5&  0.42& 0.15&  2.16& 0.09& -48.8& -49.0& 170.1& 49.1 \nl
 14&   -21.95&   64.34& 4&  0.40& 0.31&  2.20& 0.62& -49.0& -49.2&  77.8& 18.1 \nl
 15&   -26.84&   75.60& 5&  0.47& 0.07&  1.84& 0.20& -48.4& -48.4&  55.2&  9.7 \nl
 16&   -38.68&   98.35& 6&  0.01& 0.09&  2.01& 0.04& -49.6& -49.6& 102.4& 14.5 \nl
 17&  -643.08&   90.22& 6&  0.08& 0.23&  1.49& 0.08& -51.1& -51.5&  41.8&  8.8 \nl
 18&  -648.91&   91.76& 3&  2.95& 0.10&  1.82& 0.13& -53.0& -53.2&   9.7&  5.7 \nl
 19&  -650.25&   92.27& 5&  1.86& 0.34&  1.94& 0.39& -52.6& -51.7&  10.4&  3.7 \nl
 20&  -650.64&   92.11& 5&  0.73& 0.10&  1.43& 0.10& -52.6& -52.8&  35.7&  3.7 \nl
 21&  -663.46&   95.49& 6&  0.47& 0.08&  1.40& 0.06& -54.3& -54.7&  22.5&  3.2 \nl
 22&  -674.24&   96.97& 3&  3.96& 0.27&  1.11& 0.14& -55.1& -55.1&   7.7&  3.1 \nl
 23&  -673.21&   96.43& 3&  4.59& 0.30&  1.04& 0.41& -54.9& -55.1&   2.1&  1.4 \nl
 24&  -680.99&   95.14& 4&  0.07& 0.09&  1.51& 0.10& -55.7& -56.0&   6.6&  1.0 \nl
 25&  -758.57&  -48.82& 3&  6.86& 0.07& -3.12& 0.15& -59.5& -60.2&   3.5&  0.6 \nl
 26&  -826.70&  -38.99& 3&  3.40& 0.04& -0.80& 0.01& -57.2& -57.2&   7.2&  2.6 \nl
 27&  -827.14&  -38.23& 5&  3.75& 0.14& -0.53& 0.06& -57.4& -57.6&   2.6&  0.9 \nl
 28&  -831.22&  -35.39& 3&  1.26& 1.04& -0.70& 0.01& -56.8& -57.4&   3.2&  1.6 \nl
 29&  -863.41&  -12.12& 5&  2.58& 0.23& -0.03& 0.12& -58.5& -60.2&  30.0&  1.0 \nl
 30&  -942.98&    1.95& 3& -4.60& 0.34& -0.18& 0.71& -64.6& -65.0&   0.6&  0.2 \nl
 31&  -963.36&   59.78& 3& -4.01& 0.99&  4.00& 0.15& -51.9& -52.4&  32.9& 18.6 \nl
 32&  -964.18&   60.34& 3& -3.28& 0.24&  3.07& 0.64& -51.7& -51.9&  57.2& 12.9 \nl
 33&  -975.72&   54.48& 5& -5.02& 0.14&  3.01& 0.13& -59.1& -60.8&   2.0&  0.8 \nl
 34&  -976.17&   54.40& 6& -4.77& 0.37&  3.31& 0.10& -59.5& -61.2&   2.6&  0.9 \nl
 35&  -982.77&   80.99& 3& -1.44& 0.14&  1.19& 0.10& -53.6& -54.1& 283.8&121.9 \nl
 36&  -983.46&   81.71& 3& -0.06& 0.06& -3.85& 0.11& -53.4& -53.6& 658.0&230.5 \nl
 37&  -983.47&   82.56& 4& -1.14& 0.08&  0.81& 0.33& -53.0& -53.4&  70.5&  9.3 \nl
 38&  -985.41&  -24.01& 3& -3.86& 0.04&  0.82& 0.07& -53.8& -54.1&  97.7& 52.0 \nl
 39&  -988.70&   75.89& 4& -2.91& 0.09&  2.81& 0.08& -61.4& -61.8&   3.5&  0.6 \nl
 40& -1034.87&  -34.44& 3& -3.20& 0.05&  0.26& 0.13& -57.8& -58.1&   6.5&  2.1 \nl
 41& -2045.95& -133.41& 7& -1.93& 0.06&  1.36& 0.09& -45.2& -45.2&  28.7&  7.8 \nl
 42& -2231.40&   84.12& 6& -3.48& 0.16&  0.67& 0.17& -62.5& -63.1& 121.1&  4.3 \nl
\tablenotetext{a}{Origin of position offsets is $\alpha_{\rm J2000}$ = 
2$^{h}$ 27$^{m}$ 04.8362$\pm$0.0004$^{s}$,
$\delta_{\rm J2000}$ = 61$^{\circ}$ 52$^{'}$ 24.607$\pm$0.002$^{''}$.
}
\tablenotetext{b}{Motion of strongest channel relative to the reference feature 1. Errors are normalized for $\chi^2$.}
\tablenotetext{c}{Time variation of the LSR velocity in the channel with the
highest peak flux.
The velocity drift is constant for some H$_{2}$O masers, while it is
fluctuating within listed velocities for others. This is most likely
caused by blending with the intensities of blended features
changing relative to each other. Full line width for
each feature is from 0.6 to 3.4 km~s$^{-1}$.}
\tablenotetext{d}{Time variation of the peak flux.}
\enddata
\end{deluxetable}
\end{footnotesize}

\subsection{The annual parallax of W3(OH)}

Once we identified a maser feature over five or more epochs,
we modeled its path in terms of its proper motion
($\mu_{\alpha},\mu_{\delta}$) and the annual parallax ($\Pi$) by using
\begin{eqnarray*}
\Delta \alpha \cos \delta & = & \Pi f_{\alpha}(\alpha,\delta,t)
+ \mu_{\alpha} t + \alpha_{0} \\
\Delta \delta & = & \Pi f_{\delta}(\alpha,\delta,t)
+ \mu_{\delta} t + \delta_{0}
\end{eqnarray*}
where $t$ is time, $\alpha_{0}$ and $\delta_{0}$ are the positions 
of a maser feature at $t=0$, and the functions $f_{\alpha}$ and $f_{\delta}$ 
are the parallax displacements in right ascension and declination, 
respectively, given, for example, by Smart (1965).

\begin{figure}
\resizebox{\hsize}{!}{\includegraphics[angle=-90]{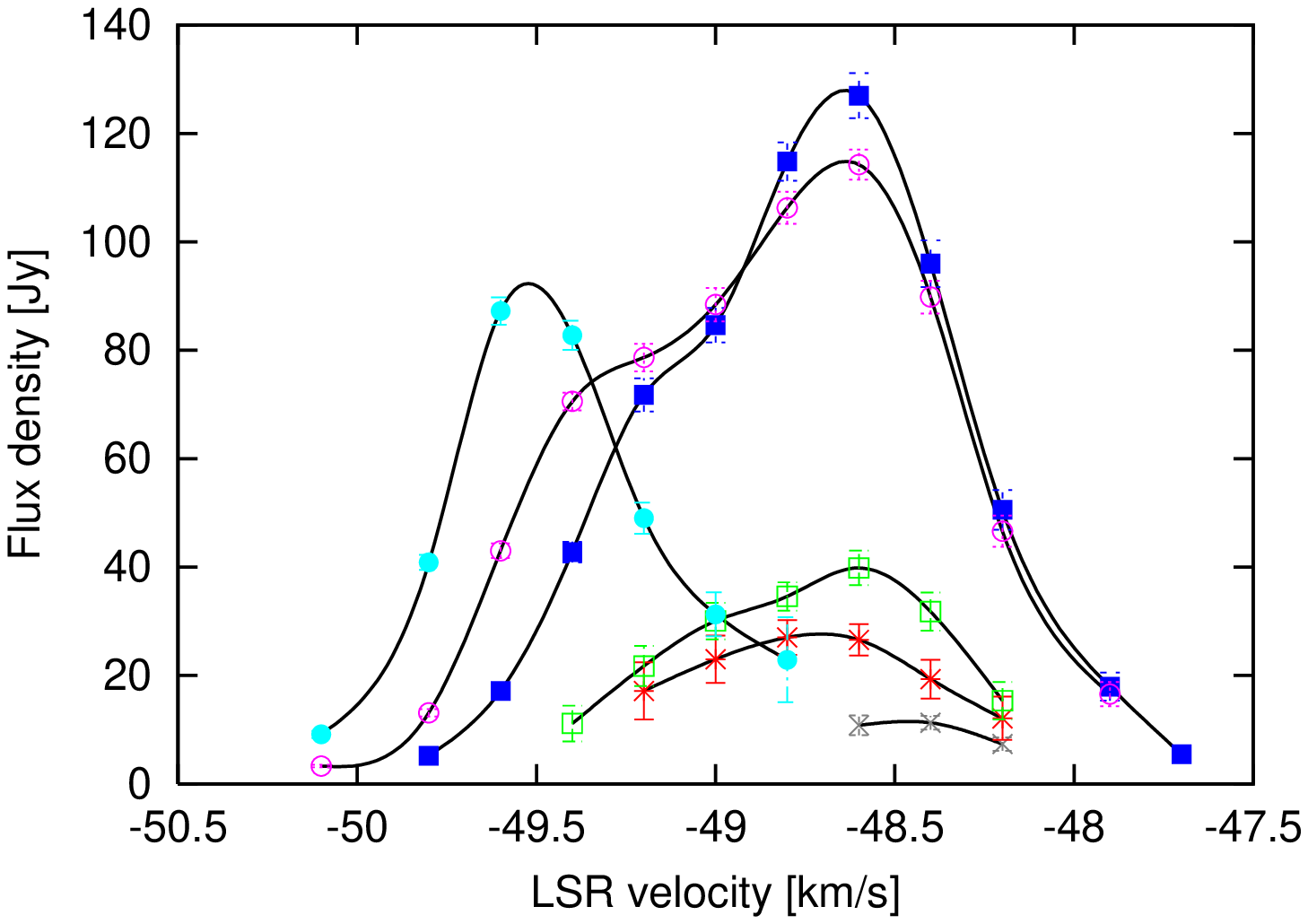}}
\caption{Spectra of component 7 at epoch 2 (exes), epoch 3 (stars), epoch 4
(open squares), epoch 5 (filled squares), epoch 6 (open circles), and
epoch 7 (filled circles).}
\label{spec5_2}
\end{figure}

First, we fitted a proper motion and an annual parallax to all velocity
channels individually.  The reduced $\chi^2$ values of the fits were very
high (10 -- 20). This was caused by unrealistically small formal errors of the
position estimates, especially for strong sources. Hence, we introduced
an error floor by adding quadratically a value of 0.05\,mas to the formal 
position error.  Possible sources of the error floor are variation of the 
centroid position of extragalactic source, residual errors in the estimation
of zenith delay corrections and blending of maser features. 
This resulted a reduced $\chi^2$ near unity. Thus, the positional accuracy of 
a single channel in a single epoch is $\sim$ 50 $\mu$as.

It is very difficult to quantify the individual error contributions. The
variations in the centroid position of the extragalactic reference source can 
be caused by unresolved structure changes and could be a few tens 
of microarcseconds per year. This motion should not influence the parallax
measurement, if the motion is constant over the time our experiment. However,
in the case of an ejection of a new jet component one could get non-linear 
motions that affect the parallax measurements. This can not be excluded since 
we did not use a second extragalactic reference source in our observations.
However, ejections of new jet components are usually accompanied by sharp rises
in the flux density. We find only small variations of the flux density of 
ICRF\,0244+624 ($\sim 10\%$) and consider this scenario as unlikely. 

Residual errors in the estimates of zenith delay corrections are much more 
likely. Since our observations were relatively short ($\sim 4$ hours) and do
not cover a large range of different elevations, it is difficult to 
separate the atmospheric and position offset contributions in the fringe phase.
These residual errors can lead to position errors of several tens of 
microarcseconds for an individual observation. Since the atmospheric 
conditions between epochs are not correlated the resulting position 
errors are random.

A major problem in the parallax fitting is that most maser features show
strong variability between the observations.  
Indeed, most features were not detected in all seven epochs. 
The flux densities of the maser
features often change by up to a factor of 10, typically causing significant
changes in the (blended) line shape and shifts in the apparent center velocity
(Fig.~\ref{spec5_2}).  
Since there is strong variability in blended spectra, one expects
some effect on the astrometric accuracy of our measurements. 
Indeed, the fits yielded a spread in parallaxes with typical values
between 0.45 and 0.55\,mas, and a few outliers at 0.4 and 0.6\,mas. This
scatter in the parallax values is much larger than the formal accuracy.
This scatter can not be explained by the two previously mentioned sources of 
error (centroid position variation of the reference source and atmospheric 
contributions), since they affect all maser components similar.

For some components, the parallax model could not match the measured
positions, as seen in reduced $\chi^2$ values larger than $\sim3$. 
(In these cases, we probably are dealing with physical changes of the source, 
such as a brightening of one part of its only marginally resolved structure.  
Also, a new component with a similar radial velocity might appear at nearly the
same position, while the existing one fades.) Other components had good fits, 
as evidenced by reduced $\chi^2$ values near unity, but still showed a large 
scatter in the parallaxes. This can be explained by a correlation between the 
parallax and proper motion parameters, coupled with position errors caused by 
structural changes in some masers. Significant correlation coefficients are 
the result of non-optimal time sampling of position measurements, caused by 
telescope scheduling and maser variability. The effect of the non-optimal 
time sampling is not identical for all maser components, because not all 
components were detected in all epochs.

\begin{figure}
\resizebox{\hsize}{!}{\includegraphics[angle=0]{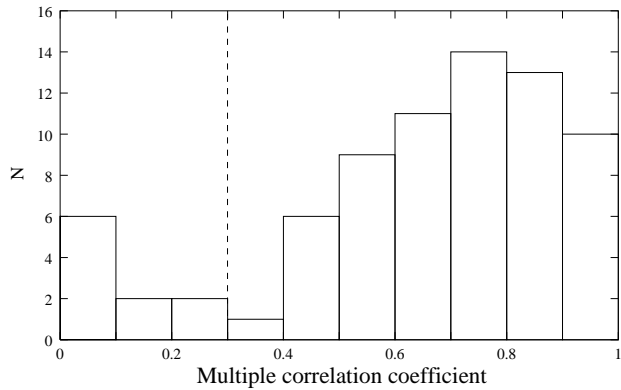}}
\caption{Histogram of the multiple correlation coefficients of the parallax
parameter for different maser features.
The dashed horizontal line marks the upper limit for the coefficient used
to determine the final parallax.}
\label{coeff}
\end{figure}

\begin{figure}
\resizebox{\hsize}{!}{\includegraphics[angle=-90]{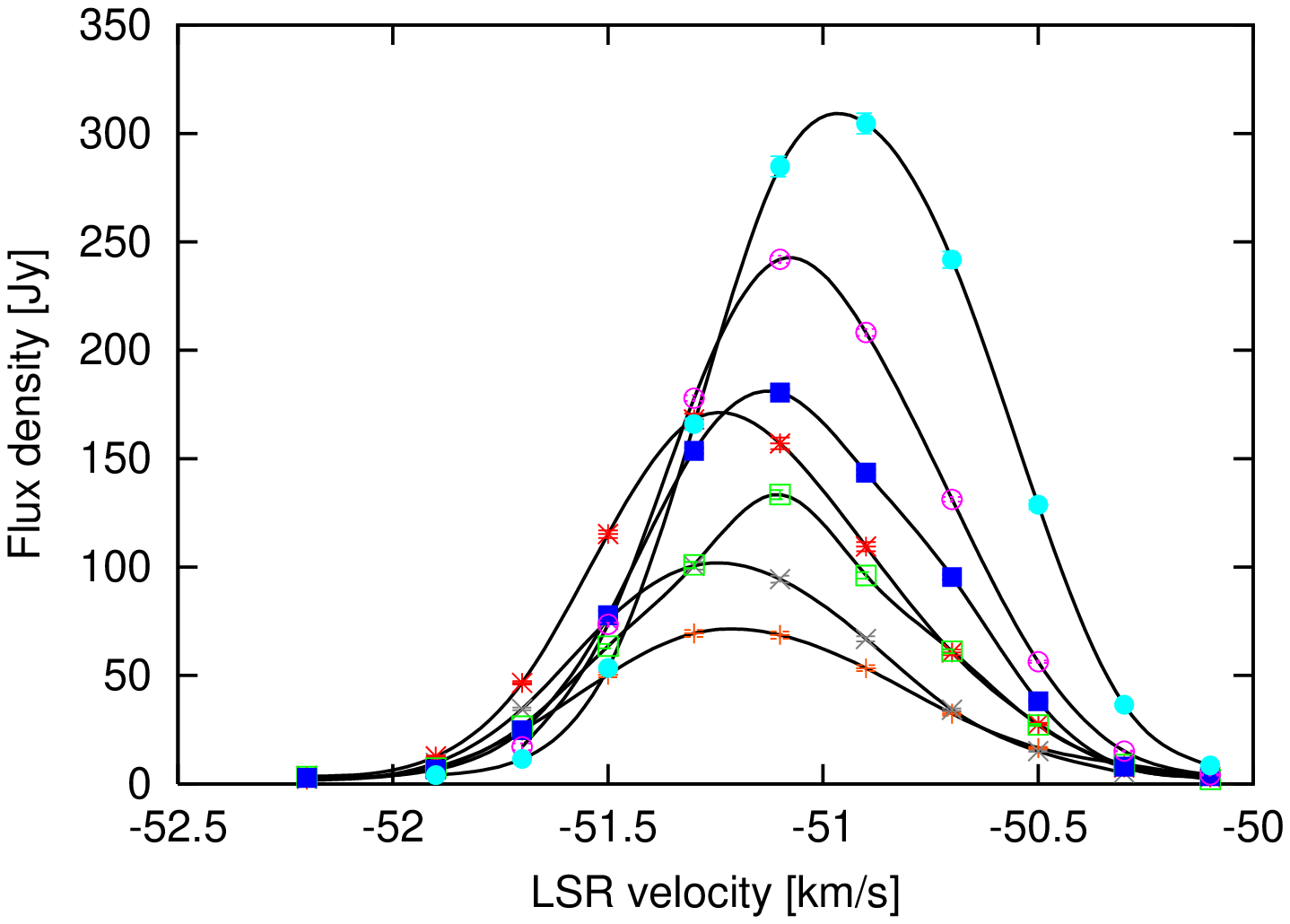}}
\resizebox{\hsize}{!}{\includegraphics[angle=-90]{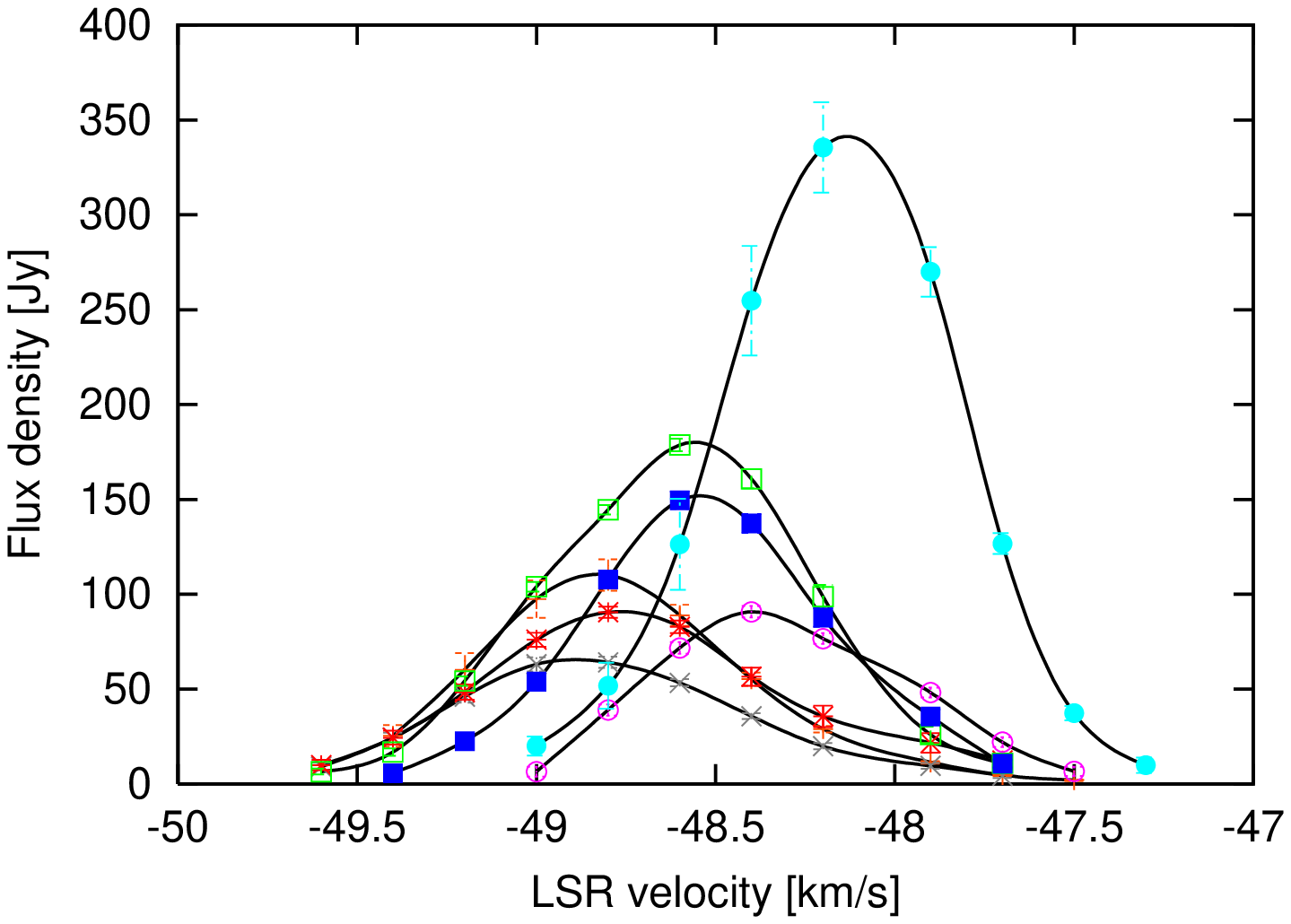}}
\caption{Spectra of component 1 (upper panel) and 4 (lower panel).
See caption of Fig.~3 for details.}
\label{spec0}
\end{figure}

  \begin{table*}
      \caption[]{Best fit annual parallax for H$_{2}$O masers in W3(OH).
The errors of the annual parallaxes are from 20 to 50 $\mu$as for
each spot, while the global fit shows 9 $\mu$as statistically.}
         \label{tab_c0}
      \[
  \begin{tabular}{p{0.18\linewidth}|cp{0.25\linewidth}cp{0.20\linewidth}}
           \hline
 Comp. & v$_{\mbox{\scriptsize LSR}}$ [km~s$^{-1}$] &$\pi$ [mas] \\
            \hline
		1		& -50.3 & 0.516 $\pm$ 0.037 \\
		1		& -50.5 & 0.479 $\pm$ 0.025 \\
		1		& -50.7 & 0.489 $\pm$ 0.021 \\
		1		& -50.9 & 0.494 $\pm$ 0.020  \\
		1		& -51.1 & 0.497 $\pm$ 0.020 \\
		1		& -51.3 & 0.490 $\pm$ 0.020 \\
		1		& -51.5 & 0.482 $\pm$ 0.018\\
            \hline
Average  &     & 0.492 $\pm$ 0.009  \\
            \hline
            \hline
		4		& -48.2 & 0.465 $\pm$ 0.047 \\
		4		& -48.4 & 0.486 $\pm$ 0.038  \\
            \hline
Average   &     & 0.476 $\pm$ 0.029  \\
            \hline
            \hline
Global fit    &     & 0.489 $\pm$ 0.009  \\
            \hline
         \end{tabular}
      \]
   \end{table*}

The fits of most features show large correlation coefficients between the 
parallax and the proper motions. A histogram of the multiple correlation
coefficient of the parallax clearly shows a bimodal distribution
(Fig.~\ref{coeff}) -- with a few values below 0.3 and a large number with
values larger than 0.4. 
Hence, we used only components that were detected in all seven epochs and
showed a multiple correlation coefficient of the parallax parameter of 
$<$ 0.3. 
This left seven channels of component 1, which is the strongest and spectrally 
broadest of all features, and two channels of component 4. 
These components also had very symmetric spatial brightness distributions.
The spectra of the two features in
all seven epochs are shown in Fig.~\ref{spec0}. The parallaxes of the
individual channel fits are in good agreement within their joint errors
(Table~\ref{tab_c0}).

\begin{figure}
\resizebox{\hsize}{!}{\includegraphics[angle=-90]{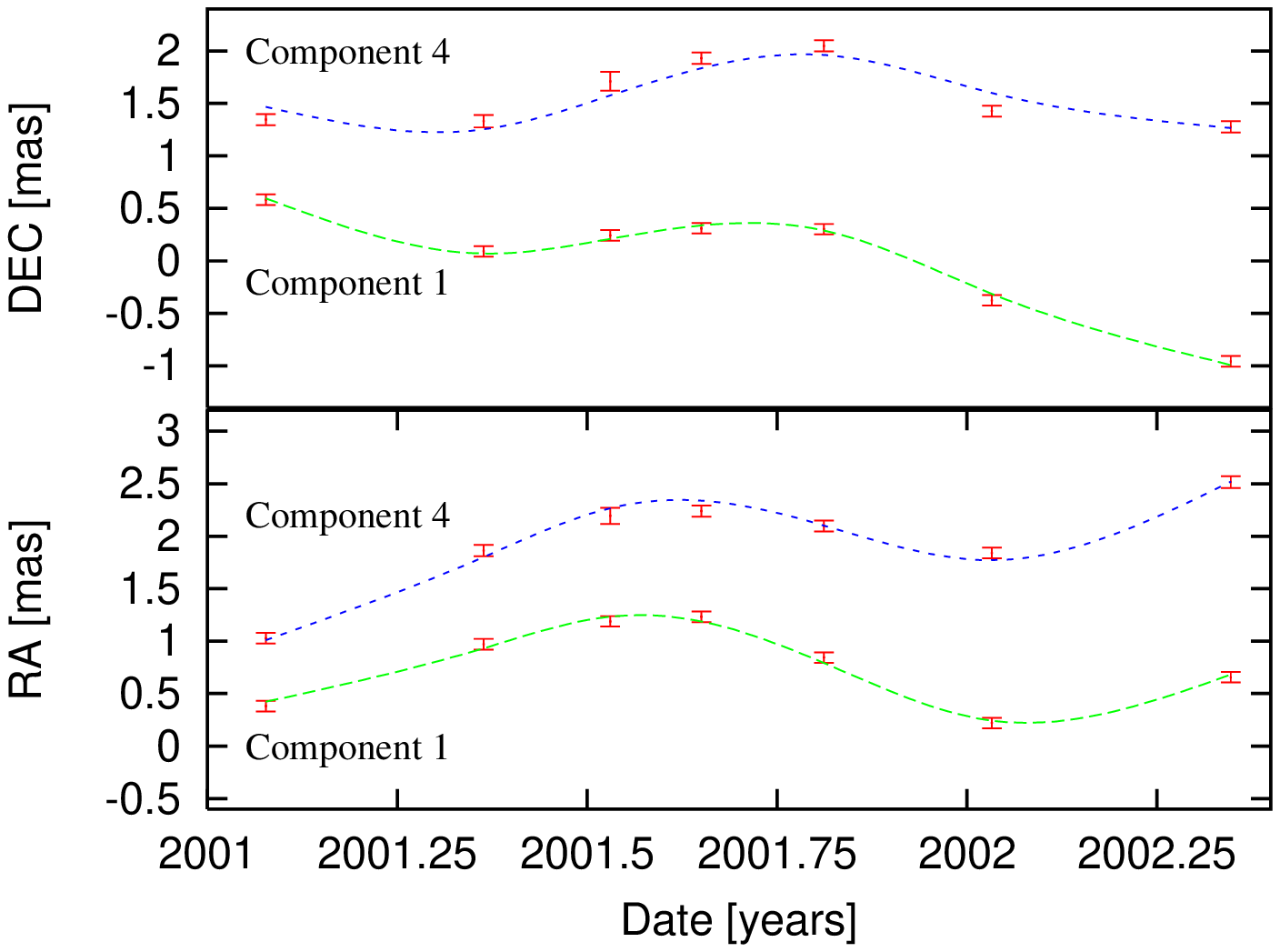}}
\caption{Positions versus time for two maser components 
(V$_{\mbox{\scriptsize LSR}}$ of --50.9 km~s$^{-1}$ for component 1 and 
--48.4 km~s$^{-1}$ for component 4) 
relative to the compact extragalactic source ICRF 0244+624. 
The top (bottom) plots show the northward (eastward) position components.
Best fit parallax and proper motions are indicated with  
dashed lines.}
\label{parallax}
\end{figure}

We also obtained a global fit to all nine channels with one parallax, but 
allowing
a different  proper motion for each channel. The measured positions and our
model for two channels are plotted in Fig.~\ref{parallax}. This global fit
yields a parallax of $0.489\pm0.009$~mas.   The 0.009~mas uncertainty is 
statistical only and does not include systematic effects.
Xu et al. (2005) see indications that parallax measurements might
show some systematic sensitivity to the angular offset of the calibrators.  
For this effect, they included a systematic parallax error of about 
0.007~mas~per degree of calibrator separation in their total uncertainty. 
This is probably caused by residual errors in the zenith delay correction.
To be conservative, since ICRF 0244+624 has a separation from W3(OH) of 
2.2 degrees, we add a systematic component of 0.015~mas to the statistical
uncertainty of 0.009~mas. Therefore, we find the annual parallax of W3(OH)-TW 
to be $0.489\pm 0.009 \pm 0.015$ mas where the first error indicates the
statistical error from the measurements while the second error describes a
systematic atmospheric contribution. 

To investigate this systematic error in more detail, we performed 
simulations where we calculated the position offsets given a parallax 
(0.5 mas), random proper motions (between -4 and +4 mas~yr$^{-1}$), position
 of W3(OH) on the sky, and dates of the observations. Then we added a random 
Gaussian error with an rms of 50 $\mu$as (the position accuracy of a single 
channel in a single epoch in our data) to each position offset, and fitted 
the simulated data set. We fitted 1000 simulated data sets and found that the 
resulting parallaxes followed a Gaussian distribution with a standard deviation
of $\sim 0.015$ mas. These errors also affect the proper motion 
fits. The difference between the true proper motion and the fitted proper 
motions in our simulations has a standard deviation of 0.04 mas~yr$^{-1}$.

The annual parallax for W3(OH)-TW corresponds to a distance of
2.04$\pm$ 0.04 $\pm$ 0.06 kpc. This is far more accurate than any previous
distance, and comparable to the result of Xu et al (2005). 
Photometric distance estimates of 2.2 kpc to an OB association near 
W3(OH) (Humphreys 1978) compare favorably with our value. However, 
the kinematic distance for a source at Galactic longitude 133.95 degrees with 
an LSR velocity near $-50$~\kms is 5~kpc. The reason for this large 
discrepancy is a peculiar motion of the W3(OH) region, 
which is discussed in detail in Xu et al. (2005).

\subsection{Relative maser motions in W3(OH)-TW}
\begin{figure*}
\centering
\includegraphics[width=12cm]{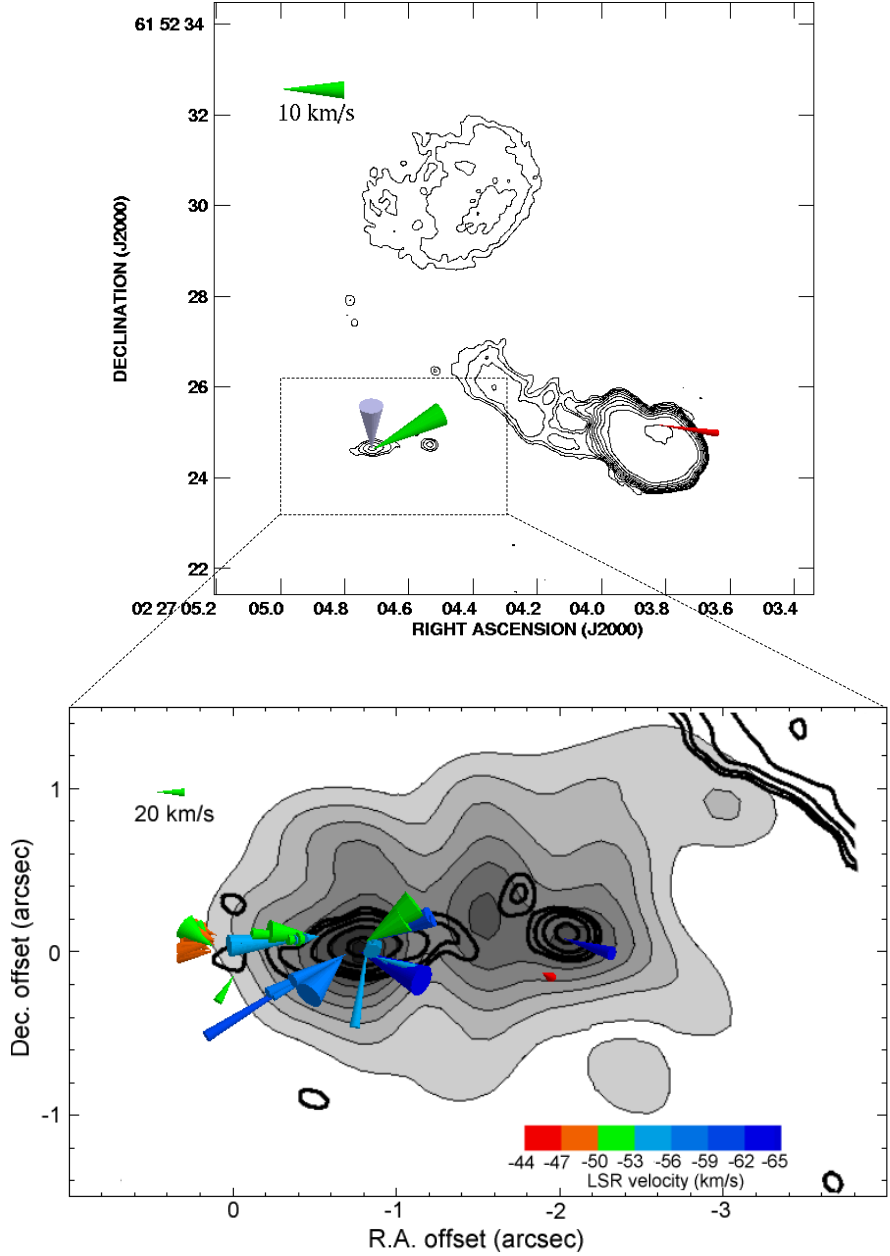}
\caption{{\it Upper panel:} the 8.4 GHz VLA continuum map of W3(OH) shown in 
contours (Wilner et al. 1999) with the absolute tangential motions of 
TW object 
and the W3(OH) UC\ion{H}{2} region indicated with cones.  The colors indicate 
radial velocities as coded in the panel at the lower right.
The gray cone shows their relative tangential motion.
{\it Lower panel:} the internal tangential motions of the W3(OH)-TW
H$_{2}$O masers.   The origin of the lower figure is 
($\alpha_{\rm J2000}$ = 2$^{h}$ 27$^{m}$ 04.8362$^{s}$,
$\delta_{\rm J2000}$ = 61$^{\circ}$ 52$^{'}$ 24.607$^{''}$).
The thin contours and grey scale represent the 220 GHz Plateau de Bure
interferometer continuum map (Wyrowski et al. 1999).  The peak positions for
the TW object in this and the 8.4 GHz map are consistent with the expansion 
center of the outflow model.
}
\label{inner-motion}
\end{figure*}

The relative proper motions of the different maser features can be described
by an outflow model. This has been successfully applied to earlier VLBI
observations of the H$_2$O masers in W3(OH) by Alcolea et al. (1993, hereafter
A93). To estimate the physical parameters of the outflow model of W3(OH), we
used the proper motions, radial velocities and positions of maser features 
(Table 2).We chose component 1 from Table~\ref{maser-position} as a 
reference feature and used its data to
re-calibrate the all of the maser data.  Component 1 was chosen because 
it is one of the strongest and least affected by blending of the masers
(see Sect.~3.1).  We used only the maser features that
were detected in at least 3 epochs. Fig.~\ref{inner-motion} (lower panel)
shows the relative internal proper motions.
We used the same method as A93, but we adopted our accurate distance of 
2.04~kpc. We estimated
the velocity and position of the center of expansion with respect to the
reference maser feature and the expansion velocity V$_{exp}$ at 1'' from the
center of expansion.  Details of the model fitting are described in A93 and
Imai et al. (2000). The best fit to the data was obtained minimizing the
expression
\begin{eqnarray*}
\chi^{2} & = & \sum
\left[
\left( \frac{u_{x}-W_{x}-v_{x}}{\sigma_{x}} \right)^{2} +
\left( \frac{u_{y}-W_{y}-v_{y}}{\sigma_{y}} \right)^{2} \right. \\
& & \qquad \left.
+ \left( \frac{u_{z}-W_{z}-v_{z}}{\sigma_{z}} \right)^{2}
\right],
\end{eqnarray*}
where $\left( u_{x}, u_{y}, u_{z} \right)$ are the 
motions of the maser feature in right ascension, declination
and the radial velocity, $W_{x}$ and $W_{y}$ are the tangential motions 
relative to the reference feature and $W_z$ is the radial velocity of the 
center of expansion. $\left( v_{x}, v_{y}, v_{z} \right)$ are components 
of a maser spot's velocity, $v$, which
is given by the equation of ${\boldmath v}$ = V$_{exp}
\left| r \right|^{\alpha} {\boldmath r} / \left| r \right|$ where
${\boldmath r} = \left( x-X_{0}, y-Y_{0}, z \right)$.
$\left( \sigma_{x},\sigma_{y}, \sigma_{z} \right)$ are components of the 
measurement uncertainty and a possible turbulent velocity of 4 km~s$^{-1}$
(added in quadrature). Finally $X_0$ and $Y_0$ are the 
positions of the center of expansion relative to the phase tracking center.
We assigned a value of $-51.0$~\kms\ for the LSR velocity of the
center of expansion (W$_{z}$), based on thermal molecular emissions 
from the location of the TW object (Wyrowski et al. 1997).

\begin{table}
\caption[]{Best fit model for H$_{2}$O maser velocity field
\label{model}}
\begin{tabular}{cr@{}c@{}lcll}
\hline
Parameter & \\
\hline
X$_{0}$ & --0&.&89  &$\pm$& 0.05  &arcsec \\
Y$_{0}$ &   0&.&00  &$\pm$& 0.03  &arcsec \\
W$_{x}$ & --10&.&0    &$\pm$& 7    &km s$^{-1}$ \\
W$_{y}$ &  16&.&0    &$\pm$& 7     &km s$^{-1}$ \\
V$_{exp}$& 13&.&0    &$\pm$& 5     &km s$^{-1}$ \\
$\alpha$ &--0&.&38  &$\pm$& 0.2  & \\
\hline
\end{tabular}
\end{table}

The results of the best model fit with a reduced $\chi^2$ of 2.7 are shown in 
Table~\ref{model}. The formal
errors were increased by $\sqrt{2.7}$ to account for the $\chi^2$.
It is not possible to compare the position of center of 
expansion between A93 and our results, since the A93 results were not 
phase-referenced. However, the parameter $\alpha$ agrees well with 
the value of A93. We find a value of $13.0\pm5$ km~s$^{-1}$ for the expansion
velocity which is slightly lower than the $20\pm2$ km~s$^{-1}$ value
obtained by A93. This discrepancy is probably not significant, since A93 
used a larger distance of 2.2 kpc that leads to larger velocities. When 
scaling the A93 results to our distance, one gets an expansion velocity of 
18.5 $\pm$ 2 km~s$^{-1}$. Then the difference is 5.5 $\pm$ 5.4  km~s$^{-1}$.

In our case, the absolute position of the reference maser feature can be
determined with respect to the extragalactic source, and transfered to
the other masers features and the center of expansion. Fig.~\ref{inner-motion}
displays the water masers and their tangential motions on
the 8.4 GHz Very Large Array (VLA) continuum map of Wilner et al. (1999) and
the 220 GHz Plateau de Bure interferometer (PdBI) continuum map of Wyrowski 
et al. (1999). Absolute positions of the TW object and the center of expansion 
of the H$_{2}$O maser outflow are listed in Table~\ref{position-tw}.
The absolute position of center of expansion is consistent with 
the TW object, in which the H$_{2}$O maser outflow originates.

\begin{table*}
\caption[]{Absolute positions of the continuum peak 
and the H$_{2}$O maser center of expansion.
\label{position-tw}}
\centering
\begin{tabular}{lll}
\hline
Position   & R.A. (J2000) & Dec.(J2000) \\
\hline
8.4 GHz & 02 27 04.713     & 61 52 24.65\\
220 GHz & 02 27 04.71       & 61 52 24.6\\
Center of expansion & 02 27 04.7103 $\pm$0.0071 & 61 52 24.607$\pm$0.030 \\
\hline
\end{tabular}
\tablenotetext{}{For the peak positions in the 8.4 and 220 GHz images
we assume uncertainties of 0.02 and 0.1 arcseconds, 
typical for high quality VLA and PdBI images.
See text for the determination of the H$_2$O center of expansion uncertainty.}
\end{table*}

\subsection{Relative motion with respect to the Ultracompact \ion{H}{2} 
region in W3(OH)}
The tangential motion of the reference feature relative to the extragalactic
background source ICRF 0244+624 obtained from the parallax/proper motion
fit is $-1.2\pm0.5$ km~s$^{-1}$ eastward and $-10.2\pm0.5$ km~s$^{-1}$ 
northward, for a distance of 2.04 kpc. Adding the tangential motion of 
the center of expansion relative to these values gives the tangential motion 
of the center of expansion: $-11.2 \pm 7 \pm 0.4$ km~s$^{-1}$
eastward and $5.8 \pm 7 \pm 0.4$ km~s$^{-1}$ northward. The second error
of 0.4 km~s$^{-1}$ comes from the systematics discussed in Sect. 3.1.
While this systematic error dominates the error in the parallax measurement,
the fitting uncertainty of the outflow model dominated the total error in the 
proper motion of the center of expansion.

On the other hand, similar astrometric phase-referencing VLBA observations
have been carried out for 12.2 GHz methanol masers associated with the
UC\ion{H}{2} region in W3(OH) (Xu et al. 2005). 
These authors have estimated the proper motion of these methanol masers
with respect to extragalactic continuum sources, and obtained a tangential
motion of $-11.1\pm0.2$ km~s$^{-1}$ and $-1.3\pm0.1$ km~s$^{-1}$, eastward and 
northward respectively.
They have not estimated the internal motions of the methanol masers.
However these are approximately 2 km~s$^{-1}$ (Moscadelli et al. 2002),
much smaller than the internal H$_{2}$O maser velocities. Thus we assign
a total uncertainty of 2 km~s$^{-1}$ to the absolute tangential motion of
the methanol masers.

Combining the two VLBA results, the center of expansion of the H$_{2}$O masers 
(presumably the TW object) moves at $-0.1\pm7.3$ km~s$^{-1}$ eastward, 
$+7.1 \pm 7.3$ km~s$^{-1}$ northward and $-7$ km~s$^{-1}$ in radial direction, 
with respect to the UC\ion{H}{2} region of W3(OH).  We used a systemic radial 
velocity of --51 km~s$^{-1}$ for the H$_{2}$O masers (TW object) and 
--44 km~s$^{-1}$ for the UC\ion{H}{2} region in W3(OH).
Converting this relative motion to a Galactic Cartesian coordinate system, 
the TW object relative to the UC\ion{H}{2} region moves toward the Galactic 
center with a velocity of $7.3\pm5.9$ km~s$^{-1}$, in the direction opposed  
to Galactic rotation with $2.1\pm4.3$ km~s$^{-1}$, and toward the North 
Galactic Pole with $6.5\pm7.3$ km~s$^{-1}$.

\section{Discussion}

\subsection{3D dynamics and structure of W3(OH)}

In the case of W3(OH), a total 
(3-dimensional) motion of the TW object with respect to the UC\ion{H}{2} 
region is $>7$ km~s$^{-1}$ (7 km~s$^{-1}$ in radial velocity plus 7 $\pm$ 10 km~s$^{-1}$ tangential motion).
What causes this large relative motion?  Assuming that
the TW object and the UC\ion{H}{2} region are gravitationally bound,
a total mass ($M_{t}$) of the W3(OH)
region can be estimated by
\begin{eqnarray*}
M_{t} &\geq& \frac{r v^{2}}{G} \\
&\geq& 1.1 \times 10^{3}
\left( \frac{r}{10^{4} \: \mbox{AU}} \right)
\left( \frac{v}{10 \: \mbox{km~s$^{-1}$}} \right)^{2} \mbox{M}_{\odot},
\end{eqnarray*}
where $r$ is a separation of W3(OH)-TW and UC\ion{H}{2} region
and $v$ is the relative motion. Since the separation along the line of sight 
is still unknown,
a separation of $10^{4}$ AU is a minimum value.
This total mass is much higher than any reasonable estimate for the combined 
mass of the stars in W3(OH)-TW the UC\ion{H}{2} region, for which we estimate 
17 and 23 M$_{\odot}$, respectively, since the spectral type of the TW object 
is approximately B0 (Wyrowski et al 1999) and that of the UC\ion{H}{2} is 
approximately O8.5 region (Harten 1976).
So, these objects do not appear to be gravitationally bound.

The only other massive stars with similarly measured 3-dimensional motions 
are the Becklin-Neugebauer object and radio source-I (related to IRc 2) in
the Orion Kleinmann-Low region.  The relative motion of these objects is very
large, $>45$ km~s$^{-1}$ (Rodriguez et al. 2005), and
Tan (2004) invokes ejection of the Becklin-Neugebauer object
from the core of the Orion Nebula Trapezium Cluster to explain this.
However, for the W3(OH) sources, given their present-day relative motions, 
a close encounter of TW and the UC\ion{H}{2} region cannot have happened.
Perhaps, one or the other (or both) had an encounter with a third object
in the recent past and are now unbound.

\subsection{Limitations of H$_{2}$O maser astrometry}

The accuracy of the present H$_{2}$O maser parallax measurements achieved
with phase-referencing VLBI is $\approx$ 10 $\mu$as.
However, the high time variability of H$_{2}$O
masers limits such measurements in significant ways.  While, we detected over
40 maser features at any epoch, and were able to trace 20 features over at 
least 5 epochs, only 2 of these 20 maser features had measurements that 
yield a reliable parallax measurement. In typical sources, few 
features might persist over a period of $\ge 1$~yr, which is optimum for 
annual parallax measurements.

If one wishes to use the parallax and proper motion results to study Galactic
structure and kinematics, one needs to model the internal motions of the 
masers. To obtain an accurate model fit of the internal motions one needs
to measure the motions of many maser features. Poor estimation of the internal 
motion (typical motions are  20 to 200 km s$^{-1}$) leads to inaccurate estimates of the 
3-dimensional motion. Of the many hundreds of known H$_{2}$O maser sources in 
the Milky Way, most will not have as many detectable components as W3(OH). 
Thus it may be difficult to study Galactic dynamics using only 
H$_{2}$O masers. 

Due to strong variability and large internal velocities, H$_2$O maser 
sources are not the best candidates to study the Galactic structure and 
dynamics. Other maser sources that show less variability and slower internal 
motions (e.g. methanol maser Xu et al. 2005) are preferable. On the other hand,
H$_{2}$O masers are much more common and also found in Galactic regions where 
no methanol masers are found. Some H$_{2}$O masers are found in the outer 
Galaxy near the edge of the
optical stellar disk (e.g. Wouterloot et al. 1988), while 
methanol masers (e.g. Pestalozzi et al. 2005) have not been found there. 
Therefore, these H$_{2}$O masers can be the best sources to measure distances 
and motions in the outer Galaxy.

\section{Conclusions}
We have measured the annual parallax of the H$_{2}$O maser source in the
W3(OH) region with phase-referenced VLBA observations. The distance of 
2.04 $\pm$ 0.07 kpc that we obtain is consistent with
previous photometric distance estimates (but with much higher accuracy)
and with the CH$_3$OH
maser parallax corresponding to 1.95 $\pm$ 0.04  kpc determined by 
Xu et al. (2005) in the related paper.

We also measured the proper motions of the W3(OH)-TW H$_{2}$O masers and
find that the TW object is moving with a speed of $>$ 7 km s$^{-1}$ with 
respect to the nearby UC\ion{H}{2} region (with its OH and CH$_3$OH 
masers).  Such a large speed difference between two massive objects
in the same star forming region is puzzling.

Although H$_2$O masers are not perfect target sources to investigate
Galactic structure and dynamics, they can still provide important information 
about regions in the Galaxy that are not accessible otherwise (e.g. the outer 
Galaxy).

\acknowledgments
KH would like to thank the Japanese VLBI group for
their assistance in preparing observations.
KH also would like to thank  J. M. Marcaide and
J. C. Guirado in the University of Val\'encia and E. Ros in MPIfR
for advice in phase-referencing VLBI astrometry.
We would like to thank F. Wyrowski in MPIfR who gave useful comments
on the TW object.
This work has been partially supported by the Spanish DGICYT grant
AYA2002-00897. KH was partially supported by the Spanish Ministerio
de Educaci\'on y Ciencia for his research.

\appendix


\begin{thebibliography}{}
\bibitem[]{Alcolea92} Alcolea, J., Menten, K. M., Moran, J. M.,
\& Reid, M. J.
1992, in Astrophysical Masers, ed. A. W. Clegg \& G. E.
Nedoluha (Heidelberg:Springer), 225
  \bibitem[]{Beasley95} Beasley, A. J., \& Conway, J. E. 1995,
Very Long Baseline Interferometry and the VLBA, ed. J. A. Zensus, P. J.
Diamond, \& P. J. Napier (San Francisco: ASP), ASP Conf. Ser., 82, 328
  \bibitem[]{Brisken02} Brisken, W. F., Benson, J. M. \& Goss, W. M.
 2002, ApJ. 571, 906
  \bibitem[]{Andreas03} Brunthaler, A., Reid, M. J. \& Falcke, H.
2005,  ASP Conf. Ser. 340: Future Directions in High Resolution Astronomy, 
p.455
\bibitem[2004]{Chatterjee04} Chatterjee, S., Cordes, J. M., Vlemmings, W. H. T., Arzoumanian, Z., Goss, W. M., Lazio, T. J. W.
2004, ApJ, 604, 339
  \bibitem[2000]{Fey00} Fey, A. L. \& Charlot, P. 2000, ApJS, 128, 17
  \bibitem[]{Harten76} Harten, R.~H.\ 1976, A\&A, 46,
109
  \bibitem[]{Humphreys78} Humphreys, R. M.
1978, ApJS, 38, 309
\bibitem[]{Imai00}Imai, H.,  Kameya, O., Sasao, T.,  Miyoshi, M.,
    Deguchi, S., Horiuchi, S. \& Asaki, Y.
2000, ApJ, 538, 751
\bibitem[]{Kurayama05} Kurayama , T., Sasao, T., Kobayashi, H.
2005, ApJ, 627, L49
  \bibitem[]{margon78} Margon, B. \& Kwitter, K. B.
1978, ApJ, 224, 43L
\bibitem[]{Moran73} Moran, J. M., Papadopoulos, G. D., Burke, B. F.,  Lo, K. Y.,
Schwartz, P. R. \& Thacker, D. L.
1973, ApJ, 185, 535
\bibitem[]{Moscadelli02} Moscadelli, L, Menten, K. M., Walmsley, C. M.
\& Reid, M. J.
2002, ApJ, 564, 813
\bibitem[]{Perryman95} Perryman, M. A. C., Lindegren, L.,
Kovalevsky, J., et al. 1995, A\&A, 304, 69
\bibitem[]{Pestalozzi05} Pestalozzi, M. R., Minier, V. \& Booth, R. S. 
2005, A\&A, 432, 737
 \bibitem[]{Reid95} Reid, M. J., Argon, A. L., Masson, C. R., Menten, K. M.,
Moran, J. M.
1995, Astrophys. J.  238, 443
\bibitem[]{Reid99} Reid, M. J., Readhead, A. C. S., Vermeulen, R. C.,
\& Treuhaft, R. N.
1999, ApJ, 524, 816
\bibitem[]{Reid04} Reid, M. J. \&  Brunthaler, A.
2004, ApJ, 616, 872
\bibitem[]{Rodriguez05} Rodriguez, L. F., Poveda, A., Lizano, S., Allen, C.
2005, ApJ, 627, L65
\bibitem[]{Ros03} Ros, E.
2005, ,  ASP Conf. Ser. 340: Future Directions in High Resolution Astronomy, 
p. 482
\bibitem[]{smart65} Smart, W.M., ``Textbook on Spherical Astronomy, Fifth 
Edition'', Cambridge Univ. Press, 1965 (Cambridge, UK), p.221
\bibitem[]{tan04} Tan, J.~C.\ 2004, ApJ, 607, L47
 \bibitem[]{Turner84} Turner, J. L. \& Welch, W. J.
1984, ApJ, 287, L81
\bibitem[]{Valdettaro01} Valdettaro, R., Palla, F., Brand, J., Cesaroni, R. 
\& Comoretto, G. et al.
2001, A\&A, 368, 845
  \bibitem[]{vLangevelde00} van Langevelde, H. J., Vlemmings, W.,
Diamond, P. J., Baudry, A. \& Beasley, A. J.
2000, A\&A, 357, 945
  \bibitem[]{Vlemmings03} Vlemmings, W., van Langevelde, H. J.,
Diamond, P. J., Habing, H. J. \& Schilizzi, R. T.
2003, A\&A. 407, 213
\bibitem[]{Wilner99} Wilner, D. J., Reid, M. J. \& Menten, K. M.
1999, ApJ,  513, 775
\bibitem[]{Wouterloot88} Wouterloot, J.G.A., Brand, J., \& Henkel, C. 
1988, A\&A, 191, 323
  \bibitem[]{wbf93}
  Wouterloot, J.G.A., Brand, J., \& Fiegle, K.
1993, A\&AS, 98, 589
\bibitem[]{Wyrowski97} Wyrowski, F., Hofner, P., Schilke, P.,
Walmsley, C. M., Wilner, D. J. \& Wink, J. E.
1997, A\&A, 320, L17
  \bibitem[]{Wyrowski99}
Wyrowski, F., Schilke, P., Walmsley, C.M., \& Menten, K.M.
1999, ApJ, 514, L43
\bibitem[]{Xu05} Xu, Y., Reid, M. J, Zheng, X. W. \& Menten, K. M.
2006, Science, 311 ,54
\end{thebibliography}
\end{document}